\documentclass[prapplied, reprint, superscriptaddress, floatfix, longbibliography]{revtex4-2}

\usepackage{mathtools}
\usepackage{textgreek}
\usepackage{graphicx}
\usepackage{bm}
\usepackage{hyperref}
\usepackage{braket}
\usepackage{xcolor}
\usepackage{multirow}
\usepackage{placeins}
\usepackage{mleftright}
\usepackage[norefs, nocites, ignoreunlbld]{refcheck}

\begin{document}


\title{Fast readout and reset of a superconducting qubit coupled to a resonator \\with an intrinsic Purcell filter}

\author{Y.~Sunada}
\affiliation{Research Center for Advanced Science and Technology (RCAST), The University of Tokyo, Meguro-ku, Tokyo 153-8904, Japan}
\author{S.~Kono}
\affiliation{RIKEN Center for Quantum Computing (RQC), Wako, Saitama 351-0198, Japan}
\author{J.~Ilves}
\affiliation{Research Center for Advanced Science and Technology (RCAST), The University of Tokyo, Meguro-ku, Tokyo 153-8904, Japan}
\author{S.~Tamate}
\affiliation{RIKEN Center for Quantum Computing (RQC), Wako, Saitama 351-0198, Japan}
\author{T.~Sugiyama}
\affiliation{Research Center for Advanced Science and Technology (RCAST), The University of Tokyo, Meguro-ku, Tokyo 153-8904, Japan}
\author{Y.~Tabuchi}
\affiliation{RIKEN Center for Quantum Computing (RQC), Wako, Saitama 351-0198, Japan}
\author{Y.~Nakamura}
\affiliation{Research Center for Advanced Science and Technology (RCAST), The University of Tokyo, Meguro-ku, Tokyo 153-8904, Japan}
\affiliation{RIKEN Center for Quantum Computing (RQC), Wako, Saitama 351-0198, Japan}

\date{April 7, 2022}

\begin{abstract}
Coupling a resonator to a superconducting qubit enables various operations on the qubit, including dispersive readout and unconditional reset.
The speed of these operations is limited by the external decay rate of the resonator.
However, increasing the decay rate also increases the rate of qubit decay via the resonator, limiting the qubit lifetime.
Here, we demonstrate that the resonator-mediated qubit decay can be suppressed by utilizing the distributed-element, multi-mode nature of the resonator.
We show that the suppression exceeds two orders of magnitude over a bandwidth of 600~MHz.
We use this ``intrinsic Purcell filter'' to demonstrate a 40-ns readout with 99.1\% fidelity and a 100-ns reset with residual excitation of less than 1.7\%.
\end{abstract}

\maketitle


\section{Introduction}

Coupling a resonator to a qubit creates a versatile experimental setup in which the qubit can be coherently manipulated and measured.
The study of this setup, known as cavity quantum electrodynamics (QED), led to demonstrations of quantum information processing using atom qubits~\cite{raimond2001colloquium}.
With the introduction of superconducting qubits, the field of circuit QED branched off and developed into one of the most promising architectures for large-scale quantum computation~\cite{blais202105circuit}.

In circuit QED, the resonator is commonly designed to be off-resonant with the qubit so that it can be used for the dispersive readout scheme~\cite{blais200406cavity}.
In this scheme, the qubit state is measured by applying a resonant probe pulse onto the resonator and detecting the phase of its reflection.
To realize a fast readout, the probe signal needs to quickly decay out of the resonator into an output transmission line for amplification and detection.
However, increasing the decay rate of the resonator also increases the rate of qubit decay via the resonator into the output line.
This shortens the qubit lifetime, degrading the fidelity of operations and measurements on the qubit.

To suppress the resonator-mediated qubit decay, various ``Purcell filters'' have been demonstrated by inserting a band-stop~\cite{reed201005fast} or band-pass~\cite{jeffrey201405fast, sete201507quantum, bronn201510broadband} filter between the resonator and the output line.
More recent proposals use an additional capacitor~\cite{bronn201510reducing} or resonator~\cite{govia201702enhanced} to cancel out the decay process.
However, adding a filter or circuit element increases the complexity and footprint of the device and is a burden when integrating a large number of qubits.

Here, we demonstrate that the resonator-mediated qubit decay can be strongly suppressed without adding any circuit elements by utilizing the distributed-element, multi-mode nature of the resonator.
Thus, we realize a minimal yet versatile circuit-QED system consisting only of a long-lived transmon qubit~\cite{koch200710chargeinsensitive} and a low-$Q$ readout resonator.
We design such a device by optimizing the position where the output line couples to the resonator.

We show that this device structure, which we call an ``intrinsic Purcell filter,'' suppresses the resonator-mediated qubit decay by more than two orders of magnitude over a bandwidth of 600~MHz.
The external decay time of the qubit into the output line is measured as $T_\mathrm{1ex} = 130$~$\mu$s, which is $\sim$30,000 times longer than the external decay time of the resonator.
Without the filter, the ratio between these decay times would have been constrained to $(\Delta / g)^2 \sim 96$, where $\Delta$ and $g$ are the qubit--resonator detuning and coupling strength, respectively~\cite{blais200406cavity}.

Taking advantage of the large contrast between the decay times of the qubit and the resonator, we demonstrate a fast, high-fidelity dispersive readout using a 40-ns probe pulse.
We obtain a readout fidelity of 99.1\% and a quantum non-demolition (QND) fidelity of 98.1\%, which are comparable with the state-of-the-art fidelities achieved using more complex systems~\cite{walter201705rapid, dassonneville202002fast}.
We also utilize the low-$Q$ resonator to demonstrate a fast unconditional reset of the first and second excited states of the transmon qubit~\cite{magnard201808fast}.
We find that the residual excitation is reduced to less than 1.7\% within 100~ns.

\section{Intrinsic Purcell filter}

\begin{figure}
\includegraphics{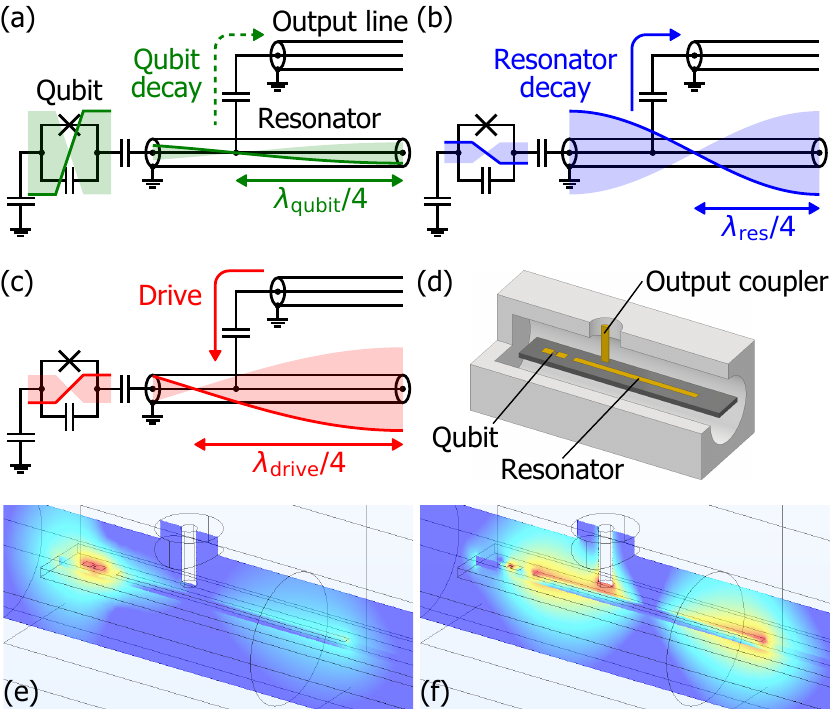}
\caption{Transmon qubit coupled to a half-wavelength resonator with an intrinsic Purcell filter. We optimize the position of the output coupler along the resonator such that the resonator-mediated qubit decay is minimized. (a)--(c)~Distributed-element circuit models with depictions of the voltage distributions of (a)~the dressed-qubit mode, (b)~the fundamental mode of the resonator, and (c)~an off-resonant drive. Here, $\lambda_\mathrm{qubit}$, $\lambda_\mathrm{res}$, and $\lambda_\mathrm{drive}$ denote the wavelengths along the resonator at each frequency. (d)~Three-quarter section illustration of the device. (e),~(f)~Finite-element electromagnetic simulations of (e)~the dressed-qubit mode and (f)~the fundamental mode of the resonator. Magnitude of the electric field on the vertical plane is visualized.}
\label{fig:modes}
\end{figure}

Figures~\ref{fig:modes}(a)--(c) show the distributed-element circuit model of our device, which consists of a transmon qubit and a half-wavelength resonator.
The resonator is a transmission line with open termination at both ends.
Conventionally, an output line couples capacitively to one end of the resonator, where the electric field is largest and strong coupling is easily realized.
Instead, we implement an intrinsic Purcell filter by optimizing the position of the output coupler along the resonator such that the resonator-mediated qubit decay is minimized.
The minimum is achieved when the distance from the open end opposite to the qubit is equal to a quarter wavelength at the qubit frequency.

The suppression of qubit decay can be understood by considering how the resonator field deforms when it dresses the qubit.
Figure~\ref{fig:modes}(a) depicts the voltage distribution of the ``dressed-qubit mode,'' which we define by approximating the transmon qubit as a linear oscillator.
The external decay of the qubit can be viewed as a result of the coupling between the dressed-qubit mode and the output line.
From this viewpoint, the decay rate depends on the voltage amplitude of the dressed-qubit mode at the position of the coupler.
Because the section of the resonator between the coupler and the open end acts as a quarter-wavelength stub, the voltage node of the dressed-qubit mode aligns with the coupler, decoupling the dressed-qubit mode from the output line.

On the other hand, the voltage node of the fundamental mode of the resonator is at the center of the resonator and does not align with the coupler, as depicted in Fig.~\ref{fig:modes}(b).
Therefore, the output line couples much more strongly to the fundamental mode of the resonator than to the dressed-qubit mode.
Furthermore, we can use the coupler to efficiently apply an off-resonant drive signal to the qubit, as depicted in Fig.~\ref{fig:modes}(c).
This is because the filter has a notch-like transmission spectrum with a stop band at the qubit frequency and therefore transmits most other frequencies.
An off-resonant drive is useful, for example, for the all-microwave reset schemes~\cite{magnard201808fast, egger201810pulsed}, one of which we demonstrate in Sec.~\ref{sec:reset}, and the all-microwave schemes for generating an itinerant microwave photon~\cite{pechal201410microwavecontrolled, zeytinoglu201504microwaveinduced, kindel201603generation}.

Alternatively, the suppression of qubit decay can be explained using the higher-harmonic modes inherent in a distributed-element resonator.
The spontaneous emission processes of the qubit mediated by each mode of the resonator are known to strongly interfere with each other~\cite{houck200808controlling}.
Since each mode has a different field distribution, we can tune its weight in the interference by changing the position of the output coupler.
At the optimal position, the total spontaneous emission rate is minimized as a result of destructive interference.

To demonstrate the simplicity of our filter, we implement it using the coaxial-transmission-line device architecture~\cite{axline201607architecture}, which is less versatile than the conventional coplanar architecture.
Figure~\ref{fig:modes}(d) shows a three-quarter section illustration of our device (see Appendix~\ref{sec:sample-and-setup} for details).
The large mode volume of a coaxial-transmission-line resonator causes it to deviate from the circuit model shown in Figs.~\ref{fig:modes}(a)--(c).
Nevertheless, we can find the optimal coupler position by calculating the field distribution of the dressed-qubit mode using a finite-element electromagnetic simulator.
The calculated electric field distributions of the dressed-qubit mode and the fundamental mode of the resonator are visualized in Figs.~\ref{fig:modes}(e) and (f).

The resonator is measured to have a resonance frequency of $\omega_\mathrm{r} / 2 \pi = 10.5106$~GHz and an external decay rate of $\kappa_\mathrm{ex} / 2 \pi = 45.7$~MHz.
The transmon is measured to have a qubit frequency of $\omega_\mathrm{eg} / 2 \pi = 8.319$~GHz and a qubit--resonator coupling strength of $g / 2 \pi = 224$~MHz.
Considering only the fundamental mode of the resonator, we can calculate the resonator-mediated decay rate of the qubit as~\cite{blais200406cavity}
\begin{equation} \label{eq:1mode-purcell}
    \Gamma'_\mathrm{ex} = \mleft( \frac{g}{\Delta} \mright)^2 \kappa_\mathrm{ex},
\end{equation}
where $\Delta \coloneqq \omega_\mathrm{eg} - \omega_\mathrm{r}$ is the qubit--resonator detuning.
The above formula predicts that the energy relaxation time $T_1$ of the qubit would be limited to $T'_\mathrm{1ex} \coloneqq 1 / \Gamma'_\mathrm{ex} = 0.33$~$\mu$s.
However, we measure our qubit to have an energy relaxation time of $T_1 = 17$~$\mu$s.
This fifty-fold enhancement of qubit lifetime demonstrates that we have successfully utilized the distributed-element, multi-mode nature of the resonator to suppress the resonator-mediated qubit decay.

To determine the maximum $T_1$ achievable in our device, we measure the resonator-mediated external decay time $T_\mathrm{1ex}$ of the qubit.
To distinguish the external decay from the internal loss of the qubit, we measure the reflection spectrum of the qubit in a continuous-wave experiment~\cite{mirhosseini201905cavity, kono202007breaking, lu202102characterizing} (see Appendix~\ref{sec:transmon-spectrum} for details).
We obtain an external decay rate $\Gamma_\mathrm{ex}$ that translates to a decay time of $T_\mathrm{1ex} \coloneqq 1 / \Gamma_\mathrm{ex} = 130$~$\mu$s.
This indicates that we have extended the lifetime limit $T_\mathrm{1ex}$ of the qubit by a factor of more than three hundred relative to the unfiltered case $T'_\mathrm{1ex}$.

\section{Transmission spectrum}

Previous works characterized the transmission spectrum of a Purcell filter by measuring the lifetime of a flux-tunable qubit at various frequencies~\cite{reed201005fast, bronn201510broadband}.
However, we cannot use this method to observe the notch in the spectrum of our filter because the qubit lifetime near the notch is overwhelmingly determined by the internal loss and not by the external decay.

Here, we take an alternative approach of driving the qubit from the output line through the filter at various drive frequencies $\omega_\mathrm{d}$.
The drive induces a Rabi oscillation or ac Stark shift of the qubit depending on whether it is resonant or off-resonant with a transition frequency of the qubit.
We use these effects to evaluate how efficiently the drive transmits through the filter and reaches the qubit.

The strength with which the qubit is driven is quantified by the amplitude of the drive Hamiltonian $\hat{\mathcal{H}}_\mathrm{d}(t) / \hbar = \Omega (\hat{b} + \hat{b}^\dagger) \cos \omega_\mathrm{d} t$, where $\hat{b}$ is the annihilation operator of the transmon qubit.
We determine the drive amplitude $\Omega$ as described below and use it to calculate the external coupling rate of the qubit to the output line as
\begin{equation} \label{eq:omega-p}
    \Gamma_\mathrm{ex}(\omega_\mathrm{d}) = \frac{\Omega^2}{4} \frac{\hbar \omega_\mathrm{d}}{P}.
\end{equation}
Here, $P$ is the drive power applied onto the device, which we calibrate using the reflection spectrum of the qubit (see Appendix~\ref{sec:drive-power} for details).
Since the external coupling rate $\Gamma_\mathrm{ex}(\omega)$ gives the external decay rate of a qubit with frequency $\omega$, it effectively gives the outward transmission spectrum of the filter.

\begin{figure}
\includegraphics{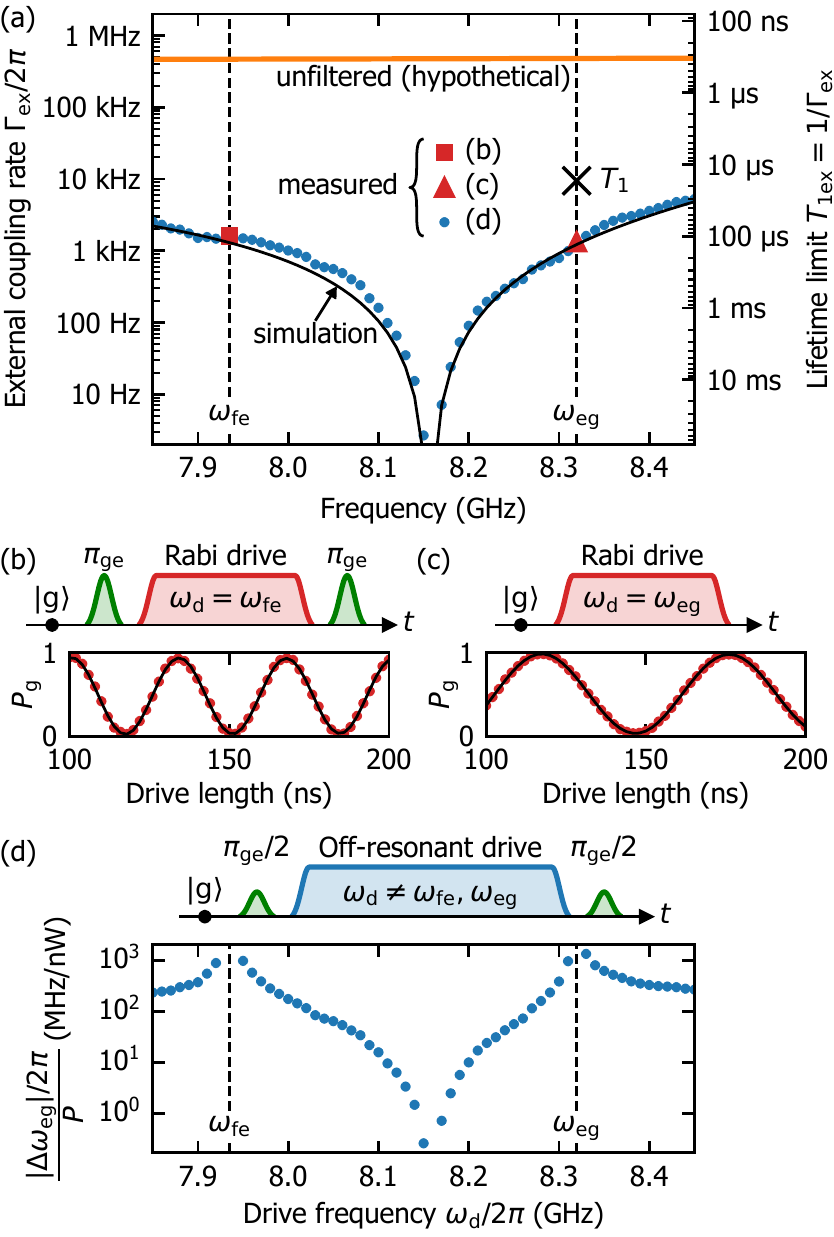}
\caption{Transmission spectrum of an intrinsic Purcell filter. (a)~External coupling rate $\Gamma_\mathrm{ex}(\omega)$ of the qubit to the output line, measured using the pulse sequences in (b)--(d). Also plotted are the result of a finite-element simulation of the device, the calculated values for a hypothetical unfiltered qubit, and the measured energy relaxation time $T_1$.
(b),~(c)~Pulse sequences and results of the $\ket{\mathrm{e}}$--$\ket{\mathrm{f}}$ and $\ket{\mathrm{g}}$--$\ket{\mathrm{e}}$ Rabi-oscillation experiments.
The measured $\ket{\mathrm{g}}$ populations $P_\mathrm{g}$ (red circles) and their fits (black lines) are plotted as functions of the drive length.
(d)~Pulse sequence and results of the ac-Stark-shift experiment.
The ratio between the ac Stark shift $\Delta \omega_\mathrm{eg}$ and the drive power $P$ is plotted as a function of the drive frequency.}
\label{fig:spectrum}
\end{figure}

Figure~\ref{fig:spectrum}(a) shows the measured external coupling rate, which agrees well with a finite-element simulation of the device (see Appendix~\ref{sec:finite-element} for details).
To evaluate the external coupling rate $\Gamma_\mathrm{ex}(\omega_\mathrm{d})$ at $\omega_\mathrm{d} = \omega_\mathrm{fe}$ and $\omega_\mathrm{eg}$, we observe the $\ket{\mathrm{e}}$--$\ket{\mathrm{f}}$ and $\ket{\mathrm{g}}$--$\ket{\mathrm{e}}$ Rabi oscillations using the pulse sequences in Figs.~\ref{fig:spectrum}(b) and (c), respectively.
Here, $\ket{\mathrm{g}}$, $\ket{\mathrm{e}}$ and $\ket{\mathrm{f}}$ denote the ground, first excited, and second excited states of the transmon qubit.
The frequencies of the Rabi oscillations correspond to $\sqrt{2} \, \Omega$ and $\Omega$, respectively.
For $\omega_\mathrm{d} \neq \omega_\mathrm{fe}, \omega_\mathrm{eg}$, we measure the drive-induced ac Stark shift of the $\ket{\mathrm{e}}$--$\ket{\mathrm{g}}$ transition frequency using the pulse sequence in Fig.~\ref{fig:spectrum}(d).
We obtain the ac Stark shift $\Delta \omega_\mathrm{eg}$ from the observed Ramsey fringes.
For each drive frequency, we choose a drive power that satisfies $\Omega \ll |\omega_\mathrm{fe} - \omega_\mathrm{d}|, |\omega_\mathrm{eg} - \omega_\mathrm{d}|$ so that we can use the perturbative formula for the ac Stark shift to calculate the drive amplitude as
\begin{equation}
    \Omega = \sqrt{\frac
        {2 (\omega_\mathrm{fe} - \omega_\mathrm{d})(\omega_\mathrm{eg} - \omega_\mathrm{d})}
        {\omega_\mathrm{fe} - \omega_\mathrm{eg}}
    \Delta \omega_\mathrm{eg}}\,.
\end{equation}

To compare the measured transmission spectrum with the unfiltered case, we use Eq.~\eqref{eq:1mode-purcell} to calculate the resonator-mediated decay rate of a qubit coupled to a hypothetical single-mode resonator.
The single-mode resonator is assumed to have the same set of parameters as the fundamental mode of our resonator.
We find that the external coupling rate of our filtered qubit is suppressed over a bandwidth of 600~MHz by more than two orders of magnitude compared to the unfiltered case.
The suppression factor exceeds one thousand for a bandwidth of 100~MHz.
Our measurement also indicates that we can achieve a lifetime limit longer than ten milliseconds by aligning the qubit frequency with the notch.

\section{Fast dispersive readout}

\begin{figure}
\includegraphics{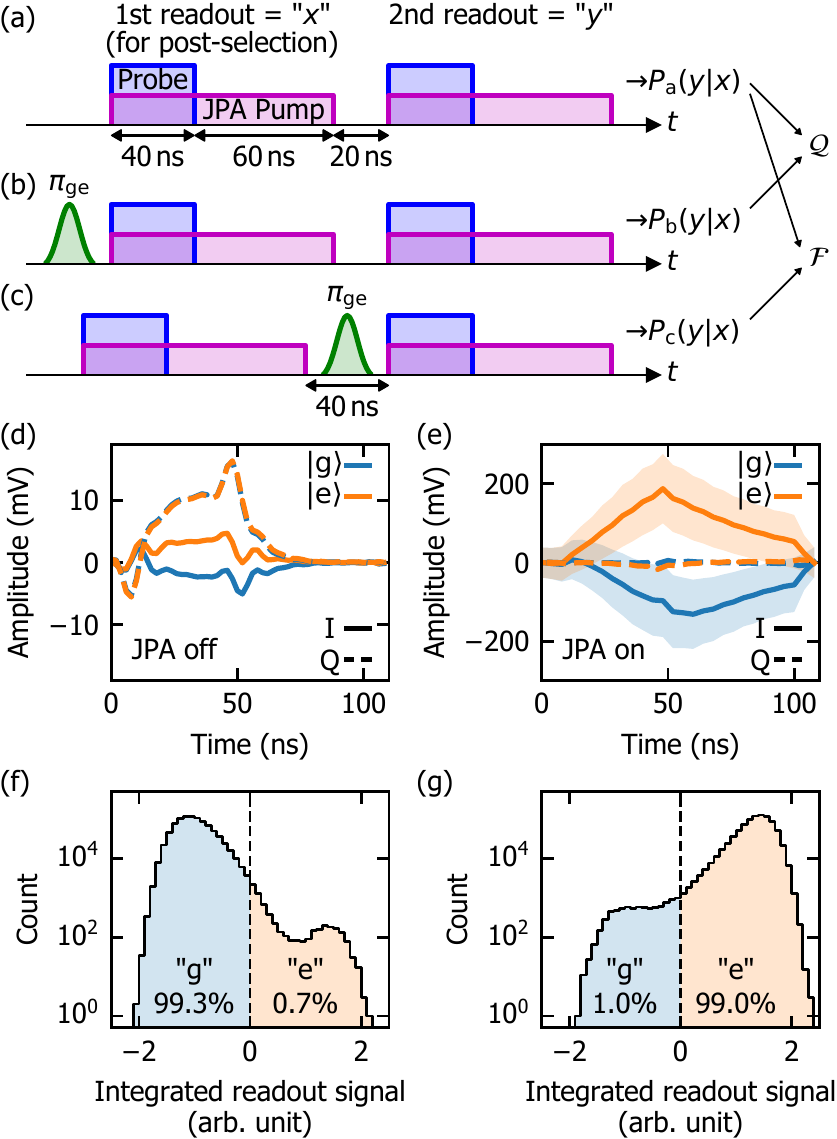}
\caption{Fast dispersive readout of a transmon qubit.
(a)--(c)~Pulse sequences used for evaluating the QND fidelity $\mathcal{Q}$ and readout fidelity $\mathcal{F}$.
(d),~(e)~Demodulated and averaged waveforms of the probe signal reflected by the readout resonator, without and with the JPA pump.
The in-phase (I) and quadrature (Q) components are defined such that the in-phase component is amplified and the quadrature component is deamplified by the JPA.
The shaded areas in (e) represent the standard deviations of the in-phase components.
(f),~(g)~Histograms of the in-phase components of the integrated readout signals for a qubit prepared in $\ket{\mathrm{g}}$ and $\ket{\mathrm{e}}$.
The dashed lines represent the threshold for discriminating the outcome of the readout.}
\label{fig:readout}
\end{figure}

Taking advantage of the low-$Q$ resonator, we demonstrate a fast dispersive readout of the transmon qubit.
As shown in the pulse sequences in Figs.~\ref{fig:readout}(a)--(c), we probe the resonator using a 40-ns square pulse.
We use a lumped-element Josephson parametric amplifier (JPA) in the phase-sensitive mode to amplify the probe signal reflected by the resonator.
We optimize the amplitude and phase of the probe pulse and the JPA pump to maximize the readout fidelity defined below.

Figure~\ref{fig:readout}(d) shows the demodulated and averaged waveforms of the reflected probe signal without amplification by the JPA.
We use the pulse sequences in Figs.~\ref{fig:readout}(a) and (c) without the JPA pump for the second readout.
The first readout is for preparing the qubit in $\ket{\mathrm{g}}$ by post-selection.
After a readout, the signal in the resonator decays by a factor of $\exp(-\kappa_\mathrm{ex} \tau) \sim 10^{-7}$ within $\tau = 60$~ns due to the large decay rate $\kappa_\mathrm{ex}$.
This rapid reset of the readout resonator allows us to quickly resume qubit operations after a readout.

Figure~\ref{fig:readout}(e) shows the probe signal after amplification by the JPA.
The signal is delayed because its bandwidth is larger than that of the JPA.
We integrate each of the collected single-shot signals after weighting it by the difference between the averaged waveforms in Fig.~\ref{fig:readout}(e).
Figures~\ref{fig:readout}(f) and (g) show the histograms of the amplified quadratures of the integrated signals for a qubit prepared in $\ket{\mathrm{g}}$ and $\ket{\mathrm{e}}$.
We fix the threshold for discriminating the outcome of the readout at zero.

Denoting by $P(y|x)$ the conditional probability that the outcome of the second readout is ``$y$'' given that the outcome of the first is ``$x$'', we obtain $P_\mathrm{a}(\mathrm{e|g}) = 0.7\%$, $P_\mathrm{b}(\mathrm{g|e}) = 3.0\%$, and $P_\mathrm{c}(\mathrm{g|g}) = 1.0\%$ for the pulse sequences in Figs.~\ref{fig:readout}(a)--(c), respectively.
Using these values, we calculate the QND fidelity as $\mathcal{Q} \coloneqq 1 - [P_\mathrm{a}(\mathrm{e|g}) + P_\mathrm{b}(\mathrm{g|e})] / 2 = 98.1\%$ and the readout fidelity as $\mathcal{F} \coloneqq 1 - [P_\mathrm{a}(\mathrm{e|g}) + P_\mathrm{c}(\mathrm{g|g})] / 2 = 99.1\%$.

We analyze the error budgets of the QND and readout infidelities using the method detailed in Appendix~\ref{sec:readout-errors}.
We find that the QND infidelity~($1 - \mathcal{Q} = 1.9\%$) consists of, in decreasing order, back action of the readout~(1.2\%), internal loss~(0.3\%), separation error~(0.3\%), and external decay~(0.1\%).
For the readout infidelity, we are only able to place upper bounds on the individual error probabilities.
We find that the readout infidelity~($1 - \mathcal{F} = 0.9\%$) consists of state-preparation error~($\le$0.6\%), back action~($\le$0.6\%), internal loss~($\le$0.3\%), separation error~(0.3\%), and external decay~($\le$0.1\%).

Since the external decay of the qubit is strongly suppressed in our device, it only adds 0.1\% to the QND and readout infidelities.
We attribute the relatively large back-action error to the strength of our probe pulse.
We estimate using measurement-induced dephasing~\cite{gambetta200610qubitphoton, kono201806quantum} that the probe power corresponds to a steady-state resonator population of 35 photons, which is 1.5 times the critical photon number $\Delta^2 / 4 g^2 = 24$~\cite{blais200406cavity}.

\section{Fast unconditional reset} \label{sec:reset}

Here, we take advantage of the low-$Q$ resonator to quickly reset the transmon qubit.
We implement an unconditional all-microwave reset protocol~\cite{magnard201808fast}, which only uses the resonator already present for dispersive readout.
In addition to resetting the first excited state $\ket{\mathrm{e}}$ into the ground state $\ket{\mathrm{g}}$, this protocol also resets the second excited state $\ket{\mathrm{f}}$, which can be populated by leakage errors during qubit operations.
In this protocol, a reset is performed by simultaneously driving the $\ket{\mathrm{f0}}$--$\ket{\mathrm{g1}}$ and $\ket{\mathrm{e0}}$--$\ket{\mathrm{f0}}$ transitions, where the numbers denote the Fock states of the resonator.
These drives transfer the excitations in $\ket{\mathrm{e0}}$ and $\ket{\mathrm{f0}}$ into $\ket{\mathrm{g1}}$, which then rapidly decay to the ground state $\ket{\mathrm{g0}}$.

Since the $\ket{\mathrm{f0}}$--$\ket{\mathrm{g1}}$ transition is a second-order process~\cite{pechal201410microwavecontrolled, zeytinoglu201504microwaveinduced}, it requires a much stronger drive signal than the $\ket{\mathrm{e0}}$--$\ket{\mathrm{f0}}$ transition.
We are able to strongly drive the $\ket{\mathrm{f0}}$--$\ket{\mathrm{g1}}$ transition without heating the experimental setup because its frequency $\omega_\mathrm{f0g1} / 2 \pi = 5.611$~GHz is outside the stopband of our filter.
The transmission of the drive signal through the filter is quantified by the external coupling rate of the qubit at the drive frequency $\Gamma_\mathrm{ex}(\omega_\mathrm{f0g1}) / 2 \pi = 11$~kHz (result of the finite-element simulation in Appendix~\ref{sec:finite-element}).
Driving through the filter is more efficient than coupling a dedicated drive line to the qubit because in the latter case the coupling rate needs to be smaller than $\Gamma_\mathrm{ex}(\omega_\mathrm{eg}) / 2 \pi = 1.3$~kHz to avoid limiting the $T_1$ of the qubit.

\begin{figure}
\includegraphics{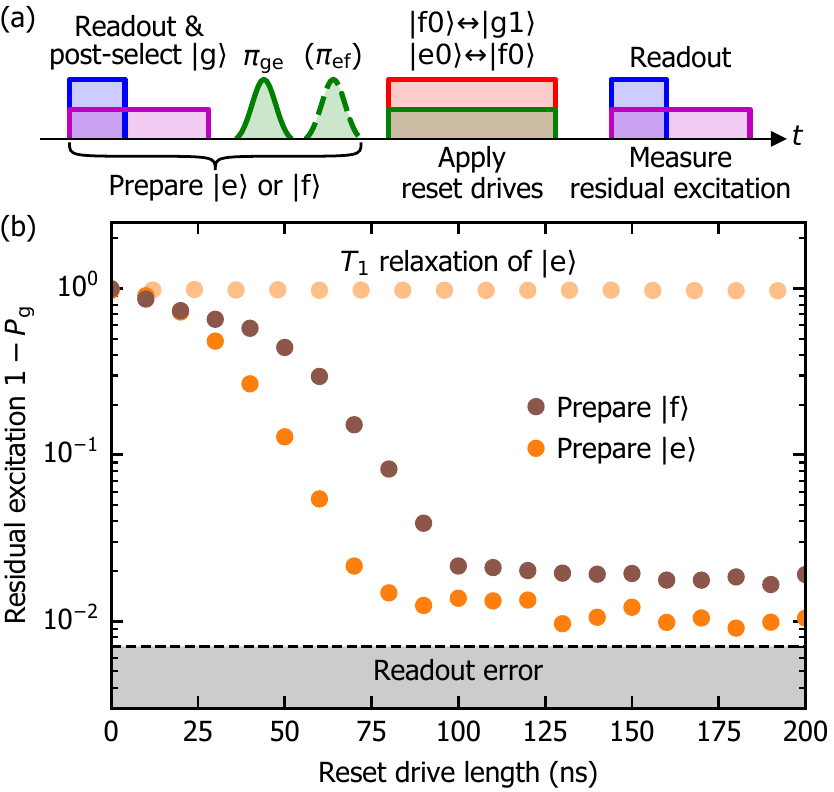}
\caption{Fast unconditional reset of the first and second excited states of a transmon qubit.
(a)~Pulse sequence for evaluating the reset.
(b)~Measured residual excitation $1 - P_\mathrm{g}$ as functions of the reset drive length.
Also plotted are the measured $T_1$ relaxation of $\ket{\mathrm{e}}$ and the upper bound for the readout error probability of detecting $\ket{\mathrm{e}}$ as ``g''.}
\label{fig:reset}
\end{figure}

To demonstrate the reset, we use the pulse sequence shown in Fig.~\ref{fig:reset}(a).
We prepare the transmon in $\ket{\mathrm{e}}$ or $\ket{\mathrm{f}}$ by first preparing a $\ket{\mathrm{g}}$ state using a post-selecting readout and then applying a $\pi_\mathrm{ge}$ pulse and optionally a $\pi_\mathrm{ef}$ pulse.
We then apply the reset drives and finally measure the residual excitation $1 - P_\mathrm{g}$ using another readout, which detects any excited state as ``e''.

Figure~\ref{fig:reset}(b) shows the measured residual excitation as functions of the reset drive length.
Accounting for the readout error of 0.5--0.7\% (obtained in Appendix~\ref{sec:readout-errors}), we find that the residual excitation is reduced to less than 1.7\% within 100~ns.
This result includes the worst case, which is when the transmon starts in the pure $\ket{\mathrm{f}}$ state.
If the transmon starts within the $\ket{\mathrm{g}}$--$\ket{\mathrm{e}}$ subspace, the residual excitation is less than 0.9\%.

We observe that the measured residual excitation remains above the upper bound of the readout error even for reset drives longer than 100~ns.
This can be explained by the leakage errors of the $\pi_\mathrm{ef}$ pulse and the $\ket{\mathrm{e0}}$--$\ket{\mathrm{f0}}$ reset drive, which may excite $\ket{\mathrm{f0}}$ into $\ket{\mathrm{h0}}$.
Here, $\ket{\mathrm{h}}$ denotes the third excited state of the transmon, which cannot be reset by this protocol.
The leakage into $\ket{\mathrm{h}}$ can also be induced by the $\ket{\mathrm{f0}}$--$\ket{\mathrm{g1}}$ reset drive because of its significant strength $\Omega / 2 \pi = 1.2$~GHz (measured following Ref.~\citenum{magnard201808fast}) relative to its detuning of $-$1.9~GHz from the $\ket{\mathrm{f}}$--$\ket{\mathrm{h}}$ transition frequency.
We expect that these leakage errors can be reduced by optimizing the waveforms of the drive pulses.

\section{Discussion}

We show that qubit decay via a readout resonator can be suppressed by shifting the position of the output coupler of the resonator.
Compared to the conventional band-pass Purcell filter~\cite{jeffrey201405fast, sete201507quantum}, this ``intrinsic Purcell filter'' offers a stronger suppression of qubit decay without introducing any additional resonance.
We demonstrate our filter using a coaxial-transmission-line resonator, which can be combined with a stub cavity to implement bosonic quantum error correction~\cite{cai202101bosonic, joshi202104quantum}.
Our filter can also be implemented using a coplanar waveguide resonator, which is favored for large-scale integration~\cite{arute201910quantum}.

Taking advantage of the low-$Q$ resonator coupled to a long-lived qubit, we demonstrate a fast dispersive readout and a fast unconditional reset of the qubit.
The fast reset also takes advantage of the band-stop nature of the filter, which allows us to efficiently apply an off-resonant drive to the qubit.
In the context of quantum error correction, the fast readout of syndrome qubits is important for minimizing the idle-time decoherence of data qubits~\cite{chen202107exponential}.
It is also important to quickly reset the leakage error in the second excited state of the qubits, which can build up and significantly degrade the error-correcting performance~\cite{fowler201310coping, mcewen202103removing}.
The low-$Q$ resonator and the efficient off-resonant driving are also useful for rapidly generating and absorbing itinerant microwave photons, which can mediate inter-node communication in a quantum network~\cite{campagne-ibarcq201805deterministic, kurpiers201806deterministic, kurpiers201910quantum, ilves202004ondemand, magnard202012microwave}.

Further investigation is needed on the effect of the notch-like filtering on $\pi$ pulses applied through the filter.
Our $\pi_\mathrm{ge}$ and $\pi_\mathrm{ef}$ pulses are calibrated to minimize phase error using the half-DRAG technique~\cite{motzoi200909simple, chow201010optimized, lucero201010reduced}.
This technique has also been shown to suppress the leakage error~\cite{chen201601measuring}, but this may not apply to the case where the drive is applied through a filter.
Because of this concern, we have intentionally detuned our qubit from the notch in the spectrum of the filter.
More advanced techniques for optimizing the shape of a $\pi$ pulse, such as those in Refs.~\citenum{machnes201804tunable} and \citenum{werninghaus202101leakage}, may be required if the qubit frequency is aligned closer to the notch.

\begin{acknowledgments}
We thank A.~van Loo and S.~Wolski for their comments on this manuscript.
This work was supported in part by UTokyo XPS, JSPS Fellowship~(Grant No.~JP21J12292), JST ERATO~(Grant No.~JPMJER1601), and MEXT Q-LEAP~(Grant No.~JPMXS0118068682).
\end{acknowledgments}

\appendix

\section{Sample and setup} \label{sec:sample-and-setup}

The transmon qubit and the inner conductor of the resonator are fabricated on a silicon substrate.
They consist of a lithographically patterned niobium film and an Al/AlO$_x$/Al Josephson junction.
The substrate is clamped inside a hole drilled in an aluminum block, which acts as the outer conductor of the resonator.
The output coupler of the resonator is a bulkhead SMA connector with a stub terminal which extends toward the inner conductor of the resonator.

\begin{figure}
\includegraphics{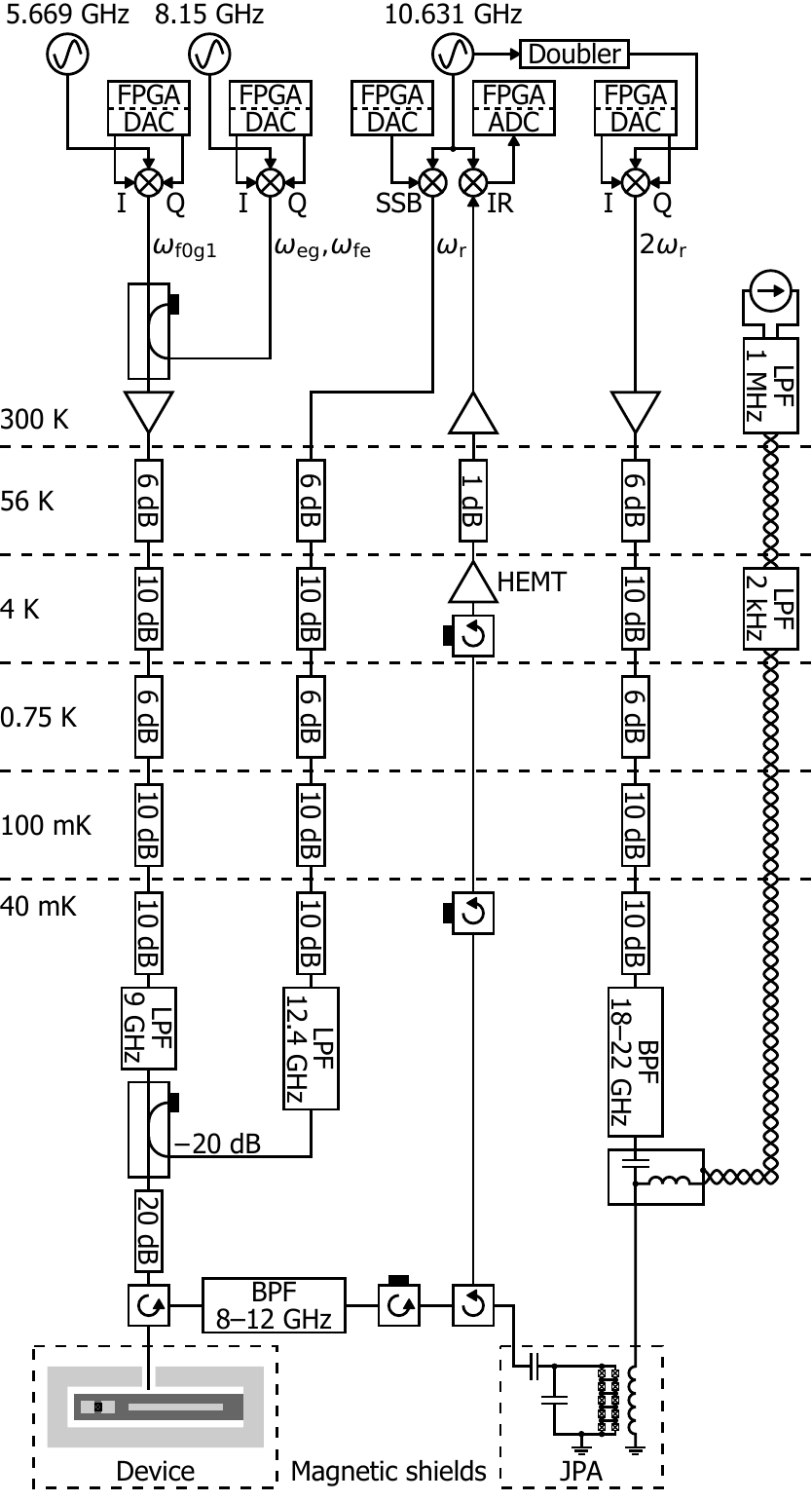}
\caption{Experimental setup.}
\label{fig:setup}
\end{figure}

\begin{table}
\caption{Measured device parameters.}
\label{tab:device-parameters}
\begin{ruledtabular}
\begin{tabular}{lcr}
$\ket{\mathrm{e}}$--$\ket{\mathrm{g}}$ transition frequency & $\omega_\mathrm{eg} / 2 \pi$ & 8.319~GHz \\
$\ket{\mathrm{f}}$--$\ket{\mathrm{e}}$ transition frequency & $\omega_\mathrm{fe} / 2 \pi$ & 7.935~GHz \\
\hline
$\ket{\mathrm{e}}$--$\ket{\mathrm{g}}$ energy relaxation time & $T_1$ & 17$\pm$1~$\mu$s \\
\multirow{2}{*}{$\ket{\mathrm{e}}$--$\ket{\mathrm{g}}$ total dephasing times $\bigg\{$} & $T_2^*$ & 5.2$\pm$0.4~$\mu$s \\
& $T_2^\mathrm{echo}$ & 15$\pm$1~$\mu$s \\
$\ket{\mathrm{f}}$--$\ket{\mathrm{e}}$ energy relaxation time & $T_\mathrm{1f}$ & 10$\pm$1~$\mu$s \\
Thermal excitation ratio & $r_\mathrm{th} \coloneqq P_\mathrm{e} / P_\mathrm{g}$ & 0.19 \\
\hline
Resonator frequency (dressed) & $\omega_\mathrm{r} / 2 \pi$ & 10.5106~GHz \\
Resonator external decay rate & $\kappa_\mathrm{ex} / 2 \pi$ & 45.7~MHz \\
Resonator dispersive shift & $2 \chi / 2 \pi$ & $-$6.9~MHz
\end{tabular}
\end{ruledtabular}
\end{table}

The device is cooled down to $\sim$40~mK in a dilution refrigerator and measured using the setup shown in Fig.~\ref{fig:setup}.
Table~\ref{tab:device-parameters} lists the measured device parameters.
We use these parameters and the perturbative formula~\cite{koch200710chargeinsensitive} to calculate the qubit--resonator coupling strength $g / 2 \pi = 224$~MHz.
The frequency, external decay rate, and dispersive shift of the resonator are measured using the method detailed in Appendix~\ref{sec:resonator-spectroscopy}.

\section{Post-selected resonator spectroscopy} \label{sec:resonator-spectroscopy}

\begin{figure}
\includegraphics{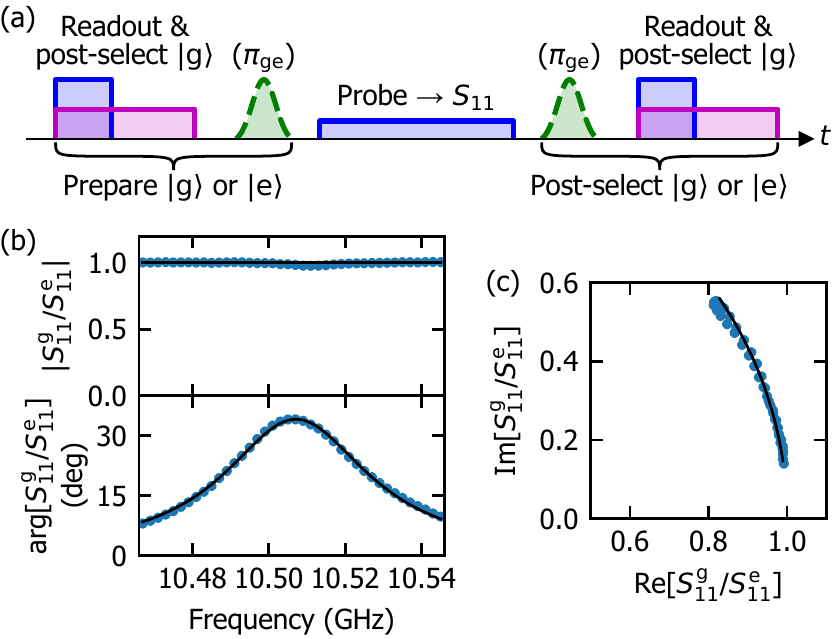}
\caption{Post-selected resonator spectroscopy. (a)~Pulse sequence. (b)~Absolute value, phase, and (c)~complex amplitude of the ratio between the measured spectra $S_{11}^\mathrm{g}(\omega) / S_{11}^\mathrm{e}(\omega)$ (blue circles) and the fit (solid black lines).}
\label{fig:resonator-spec}
\end{figure}

The qubit-state-dependent dispersive shift of the resonator could be measured by a continuous-wave spectroscopy if the two resonances corresponding to $\ket{\mathrm{g}}$ and $\ket{\mathrm{e}}$ could be resolved.
However, this is difficult for our device because the linewidth of the resonator is much larger than the dispersive shift.
Here, we use the pulse sequence in Fig.~\ref{fig:resonator-spec}(a) to prepare the qubit in $\ket{\mathrm{g}}$ or $\ket{\mathrm{e}}$ by post-selection and measure the corresponding resonator spectra $S_{11}^\mathrm{g}(\omega)$ and $S_{11}^\mathrm{e}(\omega)$.
Then, we calculate their ratio to cancel out the background fluctuation due to the microwave components in the measurement chain.

Figures~\ref{fig:resonator-spec}(b) and (c) show the ratio between the measured spectra.
We fit the data using the model function $S_{11}^\mathrm{g}(\omega) / S_{11}^\mathrm{e}(\omega)$ where
\begin{subequations}
\begin{align}
    S_{11}^\mathrm{g}(\omega) & = 1 - \frac{\kappa_\mathrm{ex}}
        {\kappa_\mathrm{ex}/2 + i(\omega - \omega_\mathrm{r})}, \\
    S_{11}^\mathrm{e}(\omega) & = 1 - \frac{\kappa_\mathrm{ex}}
        {\kappa_\mathrm{ex}/2 + i(\omega - \omega_\mathrm{r} - 2 \chi)}.
\end{align}
\end{subequations}
We assume that the internal decay rate of the resonator is negligibly small relative to the external decay rate $\kappa_\mathrm{ex}$.

\section{Reflection spectrum of the qubit} \label{sec:transmon-spectrum}

\begin{figure}
\includegraphics{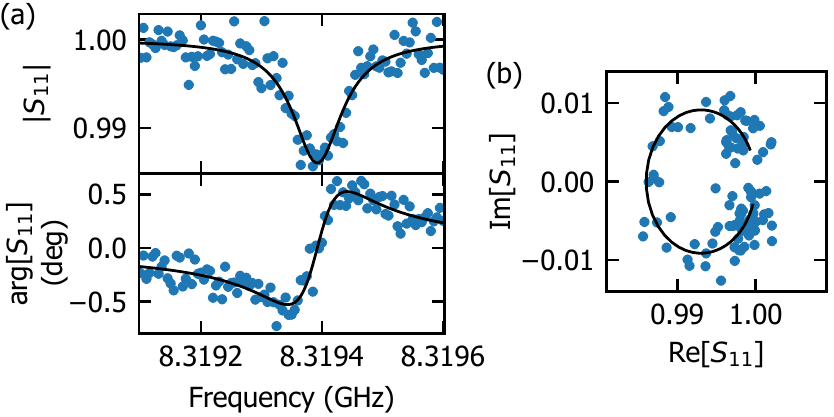}
\caption{Reflection spectrum of the qubit. (a)~Absolute value, phase, and (b)~complex amplitude of the reflection spectrum (blue circles) and the fit (black lines).}
\label{fig:qubit-s11}
\end{figure}

\begin{table}
\caption{Parameters obtained from the refection spectrum of the qubit shown in Fig.~\ref{fig:qubit-s11}.}
\label{tab:s11-fit}
\begin{ruledtabular}
\begin{tabular}{lcr}
External decay rate & $\Gamma_\mathrm{ex} / 2\pi$ & 1.33$\pm$0.09~kHz \\
Total dephasing rate & $\Gamma_2 / 2\pi$ & 38$\pm$3~kHz \\
Qubit saturation & $s$ & 0.7$\pm$0.2 \\
\end{tabular}
\end{ruledtabular}
\end{table}

We estimate the external decay rate $\Gamma_\mathrm{ex}$ of the qubit by measuring its reflection spectrum in a continuous-wave experiment~\cite{mirhosseini201905cavity, kono202007breaking, lu202102characterizing}.
Figure~\ref{fig:qubit-s11} shows the measured spectrum.
We fit the data using the model function
\begin{equation}
    S_{11}(\omega) = 1 -
        \frac{\tilde{\Gamma}_\mathrm{ex}}{\Gamma_2}
        \frac
            {1 - i (\omega_\mathrm{eg} - \omega) / \Gamma_2}
            {1 + s + (\omega_\mathrm{eg} - \omega)^2 / \Gamma_2^2},
\end{equation}
which is derived using the input--output relation~\cite{gardiner198506input} and the master equation of the qubit.
The parameters of the fit are the qubit frequency $\omega_\mathrm{eg}$, the total dephasing rate $\Gamma_2$, the qubit saturation
\begin{equation}
    s = \frac{\Omega^2}{(\Gamma_\mathrm{g \to e} + \Gamma_\mathrm{e \to g}) \Gamma_2},
\end{equation}
and the modified external decay rate
\begin{equation}
    \tilde{\Gamma}_\mathrm{ex} = \frac{1 - r_\mathrm{th}}{1 + r_\mathrm{th}} \Gamma_\mathrm{ex}.
\end{equation}
Here, $\Gamma_\mathrm{g \to e}$ and $\Gamma_\mathrm{e \to g}$ are the energy relaxation rates, and $r_\mathrm{th} \coloneqq P_\mathrm{e} / P_\mathrm{g}$ is the thermal excitation ratio, which is measured independently and listed in Table~\ref{tab:device-parameters}.
We also include as additional parameters the scaling, phase offset, and electrical delay, which are corrected for in the plots.
Table~\ref{tab:s11-fit} lists the parameters obtained from the fit.

\section{Estimation of drive power} \label{sec:drive-power}

We use the external decay rate obtained in Appendix~\ref{sec:transmon-spectrum} to estimate the drive power applied onto the device, which is required in Eq.~\eqref{eq:omega-p} to calculate the transmission spectrum of the filter.
First, we calculate the drive power $P$ used for the $\ket{\mathrm{g}}$--$\ket{\mathrm{e}}$ Rabi-oscillation experiment by substituting into Eq.~\eqref{eq:omega-p} the external decay rate $\Gamma_\mathrm{ex}$ obtained in Appendix~\ref{sec:transmon-spectrum} and the drive amplitude $\Omega$ obtained from Fig.~\ref{fig:spectrum}(c).
By comparing this drive power $P$ with the drive power going into the dilution refrigerator, we determine the attenuation of our cryogenic wiring at the qubit frequency $\omega_\mathrm{eg}$.
We then assume that the attenuation is constant within the frequency range shown in Fig.~\ref{fig:spectrum}(a).
To justify this assumption, we measure the attenuation at room temperature and find that the variation across the frequency range is less than 0.6~dB.

\section{Finite-element simulation} \label{sec:finite-element}

\begin{figure}
\includegraphics{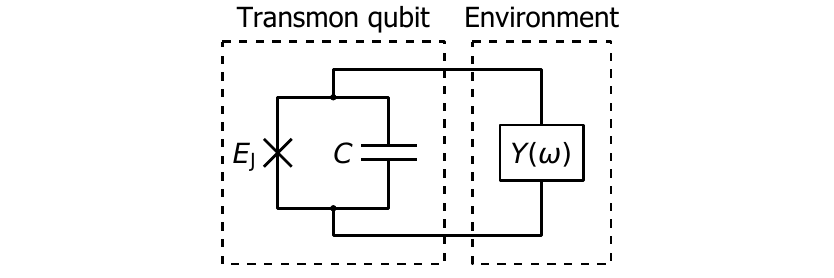}
\caption{Modeling the environment of a transmon qubit using an admittance $Y(\omega)$. $E_\mathrm{J}$ is the Josephson energy, and $C$ is the total capacitance of the transmon.}
\label{fig:admittance}
\end{figure}

\begin{figure}
\includegraphics{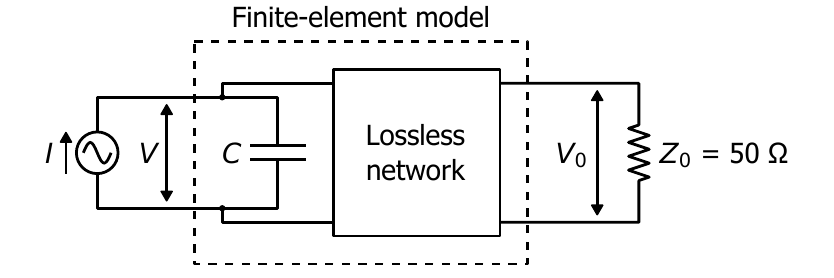}
\caption{Circuit model for the finite-element simulation of the device. The Josephson junction is replaced by an ac current source $I$, and the output line is replaced by an equivalent resistor $Z_0$.}
\label{fig:admittance-z0}
\end{figure}

Here, we explain how to calculate the external coupling rate of a qubit by finite-element electromagnetic simulation.
In the finite-element model of our device, we first fine-tune the permittivity and horizontal position of the substrate and the vertical position of the connector to match the measured resonator frequency, notch frequency, and resonator decay rate, respectively.

We calculate the external coupling rate using the real part of the admittance $Y(\omega)$ of the environment defined in Fig.~\ref{fig:admittance}.
We replace the Josephson junction by an ac current source in the finite-element model, as shown in Fig.~\ref{fig:admittance-z0}.
Then, the real part of the admittance is calculated as
\begin{equation}
    \operatorname{Re}[Y(\omega)] = \operatorname{Re}\mleft[ \frac{I(\omega)}{V(\omega)} \mright],
\end{equation}
where $I(\omega)$ and $V(\omega)$ are the current and voltage across the source.

However, this calculation is sensitive to numerical errors because the real part of $I(\omega) / V(\omega)$ is usually smaller than the imaginary part by many orders of magnitude.
This is especially problematic around the notch of our filter, where the real part approaches zero.
To avoid this issue, we use the voltage across the output port $V_0(\omega)$ to calculate the real part of the admittance as
\begin{equation}
    \operatorname{Re}[Y(\omega)] = \frac{1}{Z_0} \mleft| \frac{V_0(\omega)}{V(\omega)} \mright|^2,
\end{equation}
where $Z_0$ is the characteristic impedance of the output line.
This formula is derived using the equality between the power supplied by the current source $\operatorname{Re}[Y(\omega)] |V(\omega)|^2$ and the power consumed by the output line $|V_0(\omega)|^2 / Z_0$.

Using the real part of the admittance $\operatorname{Re}[Y(\omega)]$, we can calculate the external coupling rate of the qubit as~\cite{leggett198408quantum}
\begin{equation} \label{eq:gamma-y}
    \Gamma_\mathrm{ex}(\omega) =
        |\varphi_\mathrm{eg}|^2 \frac{\hbar \omega}{2 e^2} \operatorname{Re}[Y(\omega)],
\end{equation}
where $\varphi_\mathrm{eg} \coloneqq \bra{\mathrm{e}}\!\hat{\varphi}\!\ket{\mathrm{g}}$ is the transition matrix element of the gauge-invariant phase difference across the qubit.
For a transmon qubit, the transition matrix element is given by $\varphi_\mathrm{eg} = (2 E_\mathrm{C} / E_\mathrm{J})^{1/4}$, where $E_\mathrm{C} = e^2/2C$ is the charging energy and $E_\mathrm{J}$ is the Josephson energy.

Note that our definition of the lifetime limit $T_\mathrm{1ex}(\omega) \coloneqq 1 / \Gamma_\mathrm{ex}(\omega)$ assumes that the qubit frequency is shifted to $\omega$ but the transition matrix element $\varphi_\mathrm{eg}$, or equivalently $E_\mathrm{J} / E_\mathrm{C}$, is kept constant.
In contrast, previous works~\cite{houck200808controlling, reed201005fast, jeffrey201405fast, bronn201510broadband, bronn201510reducing} assume a fixed $E_\mathrm{C}$, which is true for a flux-tunable qubit and has the advantage that Eq.~\eqref{eq:gamma-y} simplifies to $\Gamma_\mathrm{ex}(\omega) \approx \operatorname{Re}[Y(\omega)] / C$.
We need to assume a constant $\varphi_\mathrm{eg}$ in our definition in order for Eq.~\eqref{eq:omega-p} to be valid for $\omega_\mathrm{d} \neq \omega_\mathrm{eg}$.

\section{Error budget of the readout} \label{sec:readout-errors}

Here, we analyze the error budgets of the readout and QND infidelities using the results of the experiments in Figs.~\ref{fig:readout}(a)--(c).
Unlike the method used in Refs.~\citenum{jeffrey201405fast}, \citenum{walter201705rapid}, and \citenum{dassonneville202002fast}, we do not assume that a histogram of the integrated readout signal is modeled by a mixture of Gaussian distributions.
This is an advantage when evaluating a fast readout because a short and strong probe signal can saturate the amplifier, distorting the histogram.
Since our method does not use the histogram, it is applicable to any QND readout scheme with binary outcome.

\begin{figure}
\includegraphics{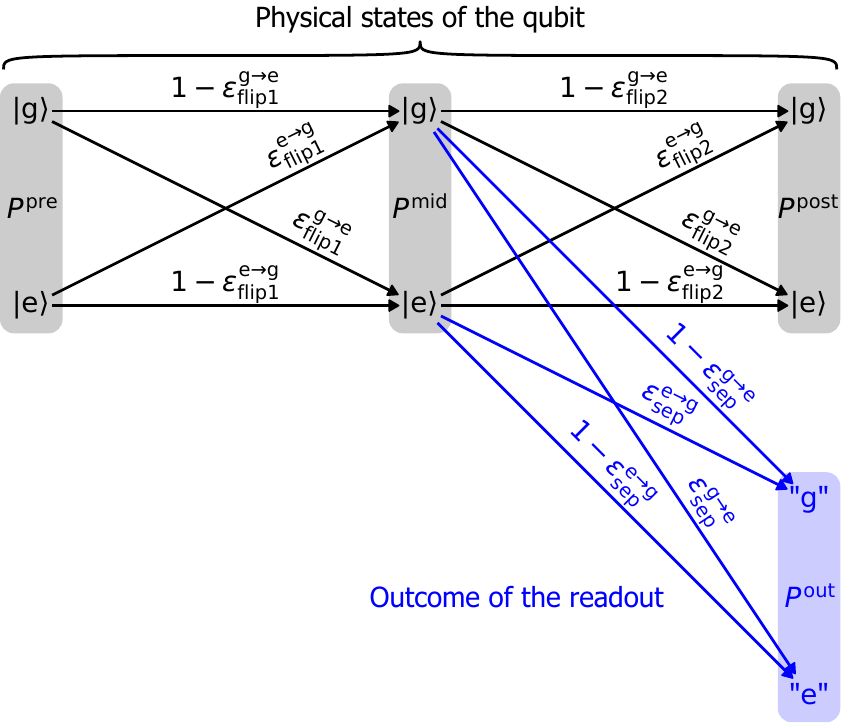}
\caption{Error processes of a single readout. Probability distributions $P^\mathrm{pre}$, $P^\mathrm{mid}$, and $P^\mathrm{post}$ represent the pre-, mid-, and post-readout states of the qubit, and $P^\mathrm{out}$ the outcome of the readout. Error probabilities $\varepsilon_\mathrm{flip1}$ and $\varepsilon_\mathrm{flip2}$ model state-flip errors, and $\varepsilon_\mathrm{sep}$ the separation error.}
\label{fig:readout-errors}
\end{figure}

\begin{figure}
\includegraphics{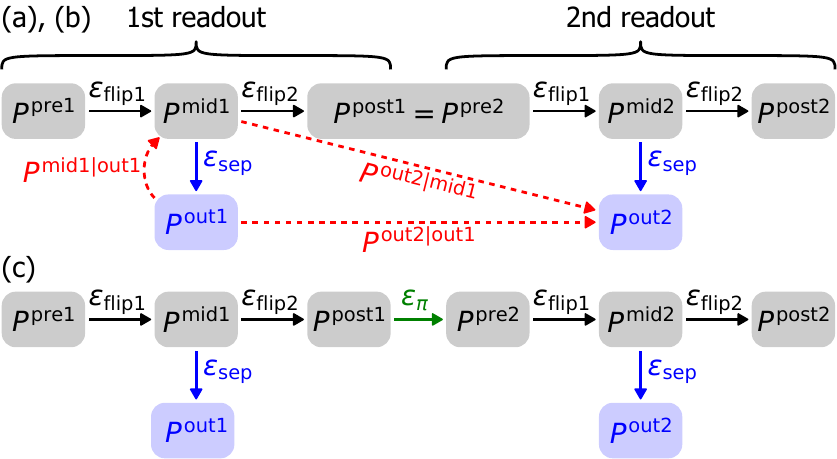}
\caption{Error processes of the experiments in Figs.~\ref{fig:readout}(a)--(c).
Here, $\varepsilon_\pi$ represents the error process of a $\pi$ pulse.
Red dashed arrows represent the expansion of Eq.~\eqref{eq:paeg-expansion}.}
\label{fig:readout-process}
\end{figure}

We model the error processes of a single readout as in Fig.~\ref{fig:readout-errors}.
Here, the state-flip errors $\varepsilon_\mathrm{flip1}$ and $\varepsilon_\mathrm{flip2}$ represent changes in the physical state of the qubit, and the separation error $\varepsilon_\mathrm{sep}$ is due to noises in the signal chain.
The early state-flip error $\varepsilon_\mathrm{flip1}$ affects both the outcome of the readout and the post-readout state, whereas the late state-flip error $\varepsilon_\mathrm{flip2}$ affects only the post-readout state.
Using this model, the experiments in Figs.~\ref{fig:readout}(a)--(c) are modeled as shown in Fig.~\ref{fig:readout-process}.

We now explain how to determine the error probabilities $\varepsilon_\mathrm{flip1}$, $\varepsilon_\mathrm{flip2}$, and $\varepsilon_\mathrm{sep}$ using the conditional probabilities measured in Figs.~\ref{fig:readout}(a)--(c).
The measured conditional probability $P(y|x) \coloneqq P^\mathrm{out2|out1}(y|x)$ is the probability distribution of the outcome of the second readout given the outcome of the first readout.
We start by expanding this into two parts as shown in Fig.~\ref{fig:readout-process}: the mid-first-readout state distribution given the first outcome $P^\mathrm{mid1|out1}$ and the second-outcome distribution given the mid-first-readout state $P^\mathrm{out2|mid1}$.
Our first goal is to eliminate the first part, which is affected by post-selection, to determine the second part, which contains just the error processes.
Applying this expansion to $P_z(\mathrm{e|g})$ for $z \in \{\mathrm{a, b}\}$, we obtain
\begin{align}
    P_z(\mathrm{e|g})
    & = P_z^\mathrm{out2|mid1}(\mathrm{e|g})
        \big[ 1 - P_z^\mathrm{mid1|out1}(\mathrm{e|g}) \big] \notag \\
    & \quad + \big[ 1 - P_z^\mathrm{out2|mid1}(\mathrm{g|e}) \big]
        P_z^\mathrm{mid1|out1}(\mathrm{e|g}). \label{eq:paeg-expansion}
\end{align}

Then, we convert the conditional probability $P_z^\mathrm{mid1|out1}(\mathrm{e|g})$ into the separation error probability $\varepsilon_\mathrm{sep}^\mathrm{e \to g} = P_z^\mathrm{out1|mid1}(\mathrm{g|e})$ using Bayes' theorem:
\begin{subequations}
\begin{align}
    P_z^\mathrm{mid1|out1}(\mathrm{e|g})
    & = \frac{P_z^\mathrm{mid1}(\mathrm{e})}{P_z^\mathrm{out1}(\mathrm{g})}
        \varepsilon_\mathrm{sep}^\mathrm{e \to g} \\
    & \approx \frac{P_z^\mathrm{out1}(\mathrm{e})}{P_z^\mathrm{out1}(\mathrm{g})}
        \varepsilon_\mathrm{sep}^\mathrm{e \to g}.
\end{align}
\end{subequations}
Here, we replace the unmeasurable mid-first-readout state distribution $P_z^\mathrm{mid1}$ by the measured first-outcome distribution $P_z^\mathrm{out1}$ using
\begin{subequations}
\begin{align}
    P_z^\mathrm{out1}(\mathrm{e})
    & = \big( 1 - \varepsilon_\mathrm{sep}^\mathrm{e \to g} \big) P_z^\mathrm{mid1}(\mathrm{e})
        + \varepsilon_\mathrm{sep}^\mathrm{g \to e} P_z^\mathrm{mid1}(\mathrm{g}) \\
    & = P_z^\mathrm{mid1}(\mathrm{e}) + O(\varepsilon_\mathrm{sep}).
\end{align}
\end{subequations}
This is a valid approximation as long as the mid-first-readout state is not too close to the ground state ($P_z^\mathrm{mid1}(\mathrm{e}) \gg \varepsilon_\mathrm{sep}$).
This is true in our device because of thermal excitation but can also be achieved by exciting the qubit into a superposition before the experiment.

Denoting the excitation ratio measured by the first readout as
\begin{equation}
    r_z \coloneqq
        \frac{P_z^\mathrm{out1}(\mathrm{e})}{P_z^\mathrm{out1}(\mathrm{g})},
\end{equation}
we can rewrite Eq.~\eqref{eq:paeg-expansion} as
\begin{multline}
    \frac{P_z(\mathrm{e|g}) - P_z^\mathrm{out2|mid1}(\mathrm{e|g})}{r_z} \\
    \approx \big[ 1 - P_z^\mathrm{out2|mid1}(\mathrm{e|g}) - P_z^\mathrm{out2|mid1}(\mathrm{g|e}) \big]
        \varepsilon_\mathrm{sep}^\mathrm{e \to g}. \label{eq:r-p-alpha}
\end{multline}
At this point, we use the fact that the conditional probabilities $P_z^\mathrm{out2|mid1}$ are equal for $z \in \{\mathrm{a, b}\}$ and denote them as $P_\mathrm{a, b}^\mathrm{out2|mid1}$.
This is because the experiments in Figs.~\ref{fig:readout}(a) and (b) are identical except for how the pre-first-readout state is prepared.
Therefore, the right hand sides of Eq.~\eqref{eq:r-p-alpha} are equal for $z \in \{\mathrm{a, b}\}$, which gives us
\begin{equation} \label{eq:equal}
    \frac{P_\mathrm{a}(\mathrm{e|g}) - P_\mathrm{a, b}^\mathrm{out2|mid1}(\mathrm{e|g})}{r_\mathrm{a}}
    \approx
    \frac{P_\mathrm{b}(\mathrm{e|g}) - P_\mathrm{a, b}^\mathrm{out2|mid1}(\mathrm{e|g})}{r_\mathrm{b}}.
\end{equation}
Solving this for $P_\mathrm{a, b}^\mathrm{out2|mid1}(\mathrm{e|g})$, we obtain
\begin{equation}
    P_\mathrm{a, b}^\mathrm{out2|mid1}(\mathrm{e|g}) \approx
    \frac{r_\mathrm{b} P_\mathrm{a}(\mathrm{e|g}) - r_\mathrm{a} P_\mathrm{b}(\mathrm{e|g})}
    {r_\mathrm{b} - r_\mathrm{a}}.
\end{equation}
Similarly, we can expand $P_z(\mathrm{g|e})$ as in Eqs.~\eqref{eq:paeg-expansion}--\eqref{eq:equal} and obtain
\begin{equation}
    P_\mathrm{a, b}^\mathrm{out2|mid1}(\mathrm{g|e}) \approx
    \frac{r_\mathrm{b} P_\mathrm{b}(\mathrm{g|e}) - r_\mathrm{a} P_\mathrm{a}(\mathrm{g|e})}
    {r_\mathrm{b} - r_\mathrm{a}}.
\end{equation}
The approximation here assumes that the mid-first-readout state is not too close to the excited state ($P_z^\mathrm{mid1}(\mathrm{g}) \gg \varepsilon_\mathrm{sep}$), which is similarly valid.
Substituting these into Eq.~\eqref{eq:r-p-alpha}, we obtain the separation error probability $\varepsilon_\mathrm{sep}^\mathrm{e \to g}$ as
\begin{equation}
    \varepsilon_\mathrm{sep}^\mathrm{e \to g}
    \approx \frac{P_\mathrm{b}(\mathrm{e|g}) - P_\mathrm{a, b}^\mathrm{out2|mid1}(\mathrm{e|g})}
    {r_\mathrm{b} \big[ 1
        - P_\mathrm{a, b}^\mathrm{out2|mid1}(\mathrm{e|g})
        - P_\mathrm{a, b}^\mathrm{out2|mid1}(\mathrm{g|e})
    \big]}
\end{equation}
and similarly
\begin{equation}
    \varepsilon_\mathrm{sep}^\mathrm{g \to e}
    \approx \frac{r_\mathrm{a} \big[
        P_\mathrm{a}(\mathrm{g|e}) - P_\mathrm{a, b}^\mathrm{out2|mid1}(\mathrm{g|e})
    \big]}{1
        - P_\mathrm{a, b}^\mathrm{out2|mid1}(\mathrm{e|g})
        - P_\mathrm{a, b}^\mathrm{out2|mid1}(\mathrm{g|e})
    }.
\end{equation}
Here, we chose $z \in \{\mathrm{a, b}\}$ appropriately to avoid subtracting nearly-equal numbers.

\begin{table}
\caption{Quantities measured by the experiments in Figs.~\ref{fig:readout}(a)--(c).}
\label{tab:measured-readout-errors}
\begin{ruledtabular}
\begin{tabular}{llcr}
\multirow{3}{*}{Fig.~\ref{fig:readout}(a)} & \multirow{2}{*}{Conditional probabilities}
& $P_\mathrm{a}(\mathrm{e|g})$ & 0.7\% \\
&& $P_\mathrm{a}(\mathrm{g|e})$ & 7.7\% \\ \cline{2-4}
& Excitation ratio & $r_\mathrm{a}$ & 0.093 \\
\hline
\multirow{3}{*}{Fig.~\ref{fig:readout}(b)} & \multirow{2}{*}{Conditional probabilities}
& $P_\mathrm{b}(\mathrm{e|g})$ & 2.2\% \\
&& $P_\mathrm{b}(\mathrm{g|e})$ & 3.0\% \\ \cline{2-4}
& Excitation ratio & $r_\mathrm{b}$ & 25 \\
\hline
\multirow{3}{*}{Fig.~\ref{fig:readout}(c)} & \multirow{2}{*}{Conditional probabilities}
& $P_\mathrm{c}(\mathrm{g|g})$ & 1.0\% \\
&& $P_\mathrm{c}(\mathrm{e|e})$ & 22.7\% \\ \cline{2-4}
& Excitation ratio & $r_\mathrm{c}$ & 0.14
\end{tabular}
\end{ruledtabular}
\end{table}

\begin{table}
\caption{Values and bounds for the individual error probabilities in a readout.}
\label{tab:error-bounds}
\begin{ruledtabular}
\begin{tabular}{lcr}
\multirow{2}{*}{Separation error}
& $\varepsilon_\mathrm{sep}^\mathrm{g \to e}$ & 0.5\% \\
& $\varepsilon_\mathrm{sep}^\mathrm{e \to g}$ & 0.1\% \\
\hline
\multirow{2}{*}{Early state-flip error}
& $\varepsilon_\mathrm{flip1}^\mathrm{g \to e}$ & $\le$0.3\% \\
& $\varepsilon_\mathrm{flip1}^\mathrm{e \to g}$ & $\le$0.9\% \\
\hline
\multirow{2}{*}{Late state-flip error}
& $\varepsilon_\mathrm{flip2}^\mathrm{g \to e}$ & $\le$0.3\% \\
& $\varepsilon_\mathrm{flip2}^\mathrm{e \to g}$ & 2.0--2.9\%
\end{tabular}
\end{ruledtabular}
\end{table}

We now go on to determine the state-flip error probabilities $\varepsilon_\mathrm{flip1}$ and $\varepsilon_\mathrm{flip2}$.
First, we break down the conditional probabilities $P_z^\mathrm{out2|mid1}$ into the individual errors:
\begin{subequations} \label{eq:error-equations}
\begin{align}
    P_\mathrm{a, b}^\mathrm{out2|mid1}(\mathrm{e|g})
        & \approx \varepsilon_\mathrm{flip2}^\mathrm{g \to e}
            + \varepsilon_\mathrm{flip1}^\mathrm{g \to e}
            + \varepsilon_\mathrm{sep}^\mathrm{g \to e}, \\
    P_\mathrm{a, b}^\mathrm{out2|mid1}(\mathrm{g|e})
        & \approx \varepsilon_\mathrm{flip2}^\mathrm{e \to g}
            + \varepsilon_\mathrm{flip1}^\mathrm{e \to g}
            + \varepsilon_\mathrm{sep}^\mathrm{e \to g}, \\
    P_\mathrm{c}^\mathrm{out2|mid1}(\mathrm{g|g})
        & \approx \varepsilon_\mathrm{flip2}^\mathrm{g \to e}
            + \varepsilon_\pi^\mathrm{g \to g}
            + \varepsilon_\mathrm{flip1}^\mathrm{e \to g}
            + \varepsilon_\mathrm{sep}^\mathrm{e \to g}, \\
    P_\mathrm{c}^\mathrm{out2|mid1}(\mathrm{e|e})
        & \approx \varepsilon_\mathrm{flip2}^\mathrm{e \to g}
            + \varepsilon_\pi^\mathrm{e \to e}
            + \varepsilon_\mathrm{flip1}^\mathrm{g \to e}
            + \varepsilon_\mathrm{sep}^\mathrm{g \to e}.
\end{align}
\end{subequations}
Here, $\varepsilon_\pi$ represents the error probabilities of a $\pi$ pulse, and we consider only the first-order errors.
We can determine $P_\mathrm{c}^\mathrm{out2|mid1}$ by expanding $P_\mathrm{c}(\mathrm{g|g})$ and $P_\mathrm{c}(\mathrm{e|e})$ in a similar manner as Eqs.~\eqref{eq:paeg-expansion}--\eqref{eq:r-p-alpha}.

Because the linear system of equations provided by Eqs.~\eqref{eq:error-equations} is not independent with respect to the four state-flip errors, we are only able to place bounds on the errors using the non-negativity of probability.
Using the measured conditional probabilities and excitation ratios listed in Table~\ref{tab:measured-readout-errors}, we obtain the values and bounds for the error probabilities listed in Table~\ref{tab:error-bounds} by linear programming.

\begin{table}
\caption{State-flip error probabilities by origin.}
\label{tab:state-flip-budget}
\begin{ruledtabular}
\begin{tabular}{lcr}
\multirow{2}{*}{Total}
& $\varepsilon_\mathrm{flip1}^\mathrm{g \to e} + \varepsilon_\mathrm{flip2}^\mathrm{g \to e}$ & 0.3\% \\
& $\varepsilon_\mathrm{flip1}^\mathrm{e \to g} + \varepsilon_\mathrm{flip2}^\mathrm{e \to g}$ & 2.9\% \\
\hline
External decay
& $\varepsilon_\mathrm{ex1}^\mathrm{e \to g} + \varepsilon_\mathrm{ex2}^\mathrm{e \to g}$ & 0.1\% \\
\hline
\multirow{2}{*}{Internal loss}
& $\varepsilon_\mathrm{in1}^\mathrm{g \to e} + \varepsilon_\mathrm{in2}^\mathrm{g \to e}$ & 0.1\% \\
& $\varepsilon_\mathrm{in1}^\mathrm{e \to g} + \varepsilon_\mathrm{in2}^\mathrm{e \to g}$ & 0.5\% \\
\hline
\multirow{2}{*}{Back action}
& $\varepsilon_\mathrm{ba1}^\mathrm{g \to e} + \varepsilon_\mathrm{ba2}^\mathrm{g \to e}$ & 0.1\% \\
& $\varepsilon_\mathrm{ba1}^\mathrm{e \to g} + \varepsilon_\mathrm{ba2}^\mathrm{e \to g}$ & 2.3\% \\
\end{tabular}
\end{ruledtabular}
\end{table}

We further break down the total state-flip error $\varepsilon_\mathrm{flip1} + \varepsilon_\mathrm{flip2}$ by origin.
We denote the error due to external decay as $\varepsilon_\mathrm{ex}$ and calculate it as $\varepsilon_\mathrm{ex1}^\mathrm{e \to g} + \varepsilon_\mathrm{ex2}^\mathrm{e \to g} = \Gamma_\mathrm{ex} \tau_\mathrm{ro}$ using the measured external decay rate $\Gamma_\mathrm{ex}$ and the duration of the readout $\tau_\mathrm{ro} = 120$~ns.
We similarly calculate the error due to internal loss $\varepsilon_\mathrm{in}$ using the measured $T_1$ and thermal excitation ratio $r_\mathrm{th}$.
The remainder is the back action of the readout $\varepsilon_\mathrm{ba}$.
Table~\ref{tab:state-flip-budget} lists these state-flip errors of different origins.

\begin{table}
\caption{Error budget of the readout infidelity.}
\label{tab:f-budget}
\begin{ruledtabular}
\begin{tabular}{lcr}
Readout infidelity &
$1 - \mathcal{F} \coloneqq [P_\mathrm{a}(\mathrm{e|g}) + P_\mathrm{c}(\mathrm{g|g})]/2$ &
0.9\% \\
\hline
\multirow{2}{*}{Preparation error} & \multicolumn{1}{l}{
$(r_\mathrm{a} \varepsilon_\mathrm{sep}^\mathrm{e \to g}
+ \varepsilon_\mathrm{flip2}^\mathrm{g \to e}
+ r_\mathrm{c} \varepsilon_\mathrm{sep}^\mathrm{e \to g}$} &
\multirow{2}{*}{$\le$0.6\%} \\
& \multicolumn{1}{r}{
${} + \varepsilon_\mathrm{flip2}^\mathrm{g \to e}
+ \varepsilon_\pi^\mathrm{g \to g}) / 2$} & \\
Back action &
$(\varepsilon_\mathrm{ba1}^\mathrm{g \to e}
+ \varepsilon_\mathrm{ba1}^\mathrm{e \to g}) / 2$ &
$\le$0.6\% \\
Internal loss &
$(\varepsilon_\mathrm{in1}^\mathrm{g \to e}
+ \varepsilon_\mathrm{in1}^\mathrm{e \to g}) / 2$ &
$\le$0.3\% \\
Separation error &
$(\varepsilon_\mathrm{sep}^\mathrm{g \to e}
+ \varepsilon_\mathrm{sep}^\mathrm{e \to g}) / 2$ &
0.3\% \\
External decay &
$\varepsilon_\mathrm{ex1}^\mathrm{e \to g} / 2$ &
$\le$0.1\% \\
\end{tabular}
\end{ruledtabular}
\end{table}

\begin{table}
\caption{Error budget of the QND infidelity.}
\label{tab:q-budget}
\begin{ruledtabular}
\begin{tabular}{lcr}
QND infidelity &
$1 - \mathcal{Q} \coloneqq [P_\mathrm{a}(\mathrm{e|g}) + P_\mathrm{b}(\mathrm{g|e})]/2$ &
1.9\% \\
\hline
Back action &
$(\varepsilon_\mathrm{ba1}^\mathrm{g \to e}
+ \varepsilon_\mathrm{ba2}^\mathrm{g \to e}
+ \varepsilon_\mathrm{ba1}^\mathrm{e \to g}
+ \varepsilon_\mathrm{ba2}^\mathrm{e \to g}) / 2$ &
1.2\% \\
Internal loss &
$(\varepsilon_\mathrm{in1}^\mathrm{g \to e}
+ \varepsilon_\mathrm{in2}^\mathrm{g \to e}
+ \varepsilon_\mathrm{in1}^\mathrm{e \to g}
+ \varepsilon_\mathrm{in2}^\mathrm{e \to g}) / 2$ &
0.3\% \\
\multirow{2}{*}{Separation error} & \multicolumn{1}{l}{
$(r_\mathrm{a} \varepsilon_\mathrm{sep}^\mathrm{e \to g}
+ \varepsilon_\mathrm{sep}^\mathrm{g \to e}
+ r_\mathrm{b}^{-1} \varepsilon_\mathrm{sep}^\mathrm{g \to e}$} &
\multirow{2}{*}{0.3\%} \\
& \multicolumn{1}{r}{
${} + \varepsilon_\mathrm{sep}^\mathrm{e \to g}) / 2$} & \\
External decay &
$(\varepsilon_\mathrm{ex1}^\mathrm{e \to g}
+ \varepsilon_\mathrm{ex2}^\mathrm{e \to g}) / 2$ &
0.1\%
\end{tabular}
\end{ruledtabular}
\end{table}

\begin{table}
\caption{Readout error probabilities.}
\label{tab:readout-errors}
\begin{ruledtabular}
\begin{tabular}{lcr}
$\ket{\mathrm{g}}$ detected as ``e'' &
$\varepsilon_\mathrm{flip1}^\mathrm{g \to e} + \varepsilon_\mathrm{sep}^\mathrm{g \to e}$ &
0.5--0.7\% \\
$\ket{\mathrm{e}}$ detected as ``g''&
$\varepsilon_\mathrm{flip1}^\mathrm{e \to g} + \varepsilon_\mathrm{sep}^\mathrm{e \to g}$ &
0.1--1.0\%
\end{tabular}
\end{ruledtabular}
\end{table}

Using the values and bounds obtained above, we can calculate the error budgets of the readout and QND infidelities as in Tables~\ref{tab:f-budget} and \ref{tab:q-budget}.
We can also calculate the readout error probabilities by excluding the state-preparation errors as in Table~\ref{tab:readout-errors}.

So far in this Appendix, we have assumed that the transmon qubit can only be in $\ket{\mathrm{g}}$ or $\ket{\mathrm{e}}$, whereas in reality it can also be in a higher-energy state $\ket{\mathrm{f}}$, $\ket{\mathrm{h}}$, etc.
Since our readout detects a higher-energy state as ``e'', post-selecting by ``e'' outcome does not guarantee that the qubit is in $\ket{\mathrm{e}}$.
This means that the conditional probabilities $P_\mathrm{a}(\mathrm{g|e})$ and $P_\mathrm{c}(\mathrm{e|e})$ may be inaccurate because the post-selected ``e'' state contains an $\ket{\mathrm{f}}$ population on the order of the thermal excitation ratio.
However, this does not affect our main claim since these quantities do not appear in the definitions of the readout and QND fidelities.


\FloatBarrier
\nocite{*}
\bibliography{bibliography}  


\end{document}